\documentclass[a4paper,12pt]{iopart}
\usepackage{iopams}
\usepackage{setstack}
\usepackage{graphicx}
\usepackage{bm}
\usepackage{epsfig}

\def\potential{W}
\def\omeg{\bar{\omega}}
\def\const{\tilde{c}}

\def\vacl{\langle 0|}
\def\vacr{|0\rangle}
\def\f{ F_{\scriptscriptstyle\mathrm{ext}}}

\newcommand{\eins}{\leavevmode\hbox{\small1\kern-3.8pt\normalsize1}}
\eqnobysec

\begin{document}

\newtheorem{definition}{Definition}[section]
\newtheorem{assumption}[definition]{Assumption}
\newtheorem{theorem}[definition]{Theorem}
\newtheorem{lemma}[definition]{Lemma}
\newtheorem{corollary}[definition]{Corollary}

\title[Collectivity in an Integrable Setting]
{Collective versus Single--Particle Motion in Quantum Many--Body Systems 
from the Perspective of an Integrable Model}
\author{Jens H\"ammerling$^\dagger$, Boris Gutkin,  Thomas Guhr }
\address{Universit\"at Duisburg-Essen, Lotharstra\ss e 1, 47048 Duisburg, Germany}
\eads{$^\dagger$ \mailto{jens.haemmerling@uni-due.de}}

\date{\today}

\begin{abstract}
  We study the emergence of collective dynamics in the integrable
  Hamiltonian system of two finite ensembles of coupled harmonic
  oscillators. After identification of a collective degree of freedom,
  the Hamiltonian is mapped onto a model of Caldeira-Leggett type,
  where the collective coordinate is coupled to an internal bath of
  phonons. In contrast to the usual Caldeira-Leggett model, the bath in
  the present case is part of the system. We derive an equation of
  motion for the collective coordinate which takes the form of a
  damped harmonic oscillator. We show that the distribution of quantum
  transition strengths induced by the collective mode is
  determined by its classical dynamics.
\end{abstract}

\maketitle

\vskip 2.0 cm
\section{Introduction}

Many-body systems show incoherent, single-particle-motion, as well as
coherent collective motion.  Historically this phenomenon received
much attention in nuclear physics where there is a wealth of data
providing information on the coexistence of collective excitations,
such as the Giant Dipole Resonance (GDR), and single particle
excitations \cite{Bohr}.  There is also strong experimental
\cite{Oos99} and theoretical \cite{Tor04} evidence that similar
effects occur in fermionic systems different from atomic nuclei.
Other examples for collective motion are vortex-generating rotations
and oscillations in Bose-Einstein condensates
\cite{But99,Mad00,Mar00}.  Furthermore collective behavior can also be
present in confined systems such as quantum dots \cite{Cle05,Yan00}.

Coherent, collective motion emerges out of incoherent, single-particle
motion whenever favored by energy conditions.  Statistical analysis of
spectra in nuclei indicates that chaotic fluctuations are due to
single-particle motion, while collective motion is predominantly
regular, for a review see Ref.~\cite{GMGW} and more recent results in
Refs~\cite{End,End2}. This generic occurrence and the coexistence of
the two forms of motion pose a fundamental challenge.  Strictly
speaking, in a generic many-body system there is not an {\it a priori}
separation of the collective motion from the single-particle dynamics.
Taking the three-dimensional Boltzmann gas with hard-wall interactions
as an example, one observes that the dynamics in the phase space of
the system is completely chaotic \cite{Arnold}. Still, we know that
the system exhibits regular collective motion in the form of sound
waves. The deep and fascinating question in this context is therefore
to understand from first principles how the regular motion emerges out
of the full phase space chaos \cite{Papenbrock}.

Whenever collective dynamics arises on the classical level one might
expect on the basis of quantum-classical correspondence that this
phenomena should be reflected in the spectral properties of the
corresponding quantum many-body Hamiltonian. One way to probe the
existence of collective excitations is to couple the system to a weak
external periodic potential $V(X)\exp (i\omega t)$ depending on a
collective mode $X$.  The presence of a collective excitation can then
be usually registered as a spike at certain energies in the
distribution of the transition strengths $T(E_n)$ between the ground
and other states of the system.  Such a large peak can be observed,
for instance, in the cross section of electric dipole radiation in
atomic nuclei at high excitation energies, when the GDR is excited.
On a phenomenological level one can obtain such a distribution of the
transition strengths from a doorway-type of Hamiltonian
\cite{Bohr,Sokolov}:
\begin{equation}{\label{doorway}}
H = \sum_{n_{c}=1}^{N_0} E_{n_{c}} |n_{c}\rangle \langle n_{c}|
         + \sum_{n,m = 1}^{N} {H}_{nm} |n\rangle \langle m|
         + \sum_{n_{c},n} {V}_{nn_{c}} |n_{c}\rangle \langle n| + c.c. \ .
\end{equation}
Here, the first term describes $N_0$ collective states $|n_{c}\rangle$
with energies $E_{n_{c}}$, the second term describes the environment
of single particle states $|n\rangle$ with $H_{nm}$ typically modeled
by a random matrix. The last term models the interaction
${V}_{nn_{c}}$ between collective and single-particle excitations.
The collective states act as doorways into the other levels of the
system.  A recent discussion can be found in Ref.~\cite{GRev}.
Although successful in the qualitative description of collective
excitation in nuclei, this model does not provide any explanation of
the physical reasons that lead to the collective behavior. We notice
that the collective and single-particle excitations are separated here
from the start, while collectivity is in fact an emergent phenomena.

Having a classical Hamiltonian whose dynamics exhibits collective
motion, what can be stated about the distribution of the transition
strengths $T(E_n)$ for the corresponding quantum problem? In
particular, it makes sense to ask under what conditions it is possible
to use models like (\ref{doorway}) and how the parameters there are
related to the classical problem. It is also of considerable interest
to understand the role of chaos in this context \cite{Brack}. Unfortunately, at
present we are lacking a genuine ``semiclassical theory" for the
emergence of collective excitations which would allow us to tackle the
problem starting from the corresponding classical dynamics. The main
goal of the present paper is to provide answers to some of the
questions posed above in the framework of a simple integrable model of
linearly coupled harmonic oscillators.  The integrability of the
system simplifies the treatment immensely. It allows for a clear
identification of a collective coordinate $X$ and an investigation of
its dynamical evolution employing an analogy with the Caldeira-Leggett
model \cite{Caldeira:1982iu}. After we fix the collective coordinate
the remaining degrees of freedom are considered as a bath which is
internal, not external as in standard models of the
Caldeira-Leggett-type \cite{Kohler1,Kohler2,Lutz,Grabert}. As a
result, it turns out that the time evolution of $X(t)$ is fully
governed by the equation of motion for a damped harmonic oscillator of
some frequency $\Omega_0$ determined by the parameters of the
many-body Hamiltonian. After this we show that under certain
conditions on the Hamiltonian of the system the averaged distribution
of $T(E_n)$ is directly connected to the corresponding classical
problem for time evolution of $X(t)$. In particular, the distribution
of the transition strengths $T(E_n)$ exhibit spikes at energies $E_n$
which are close to the energies $E_n = E_0 +n\hbar\Omega_0$ --- where
$E_0$ is the ground state energy --- of the collective oscillations,
while the width of these spikes is controlled by the classical decay
rate $\gamma$ of these oscillations.  Even though the considered model
does not involve chaotic features it serves as a testing ground to
address the emergence of collective dynamics in a many-body system.
Furthermore, it allows to see the effect of the absence of dynamical
chaos on the distribution of $T(E_n)$ and set up the ground for future
investigations.
 
The article is organized as follows. In Sec. \ref{Sec1} we introduce
our model and map it to a Caldeira-Leggett-like system. In order to
illustrate the general procedure we treat the special configuration of
two simple coupled chains in Sec. \ref{Sec3}. In Sec. \ref{Sec4} we
derive the equation of motion for the collective coordinate and obtain
an expression for the spectral density which encodes the crucial
physical properties of our model.  In Sec. \ref{Sec6} we investigate
the distribution of transition strengths between the ground state and
excited states and relate the result with the dynamics of collective
motion.

\section{Coupled chains of oscillators}{\label{Sec1}}

In Sec. \ref{Sec21} we define the model. After defining a collective
coordinate we map the system onto a Caldeira-Leggett-like model in
Sec. \ref{collectivecoordinate}.

\subsection{The model}{\label{Sec21}}

We consider two identical chains of one-dimensional coupled harmonic oscillators each
consisting of $N$ particles with positions $x_j$, $j=1\dots N$ and
momenta $p_j$, $j=1\dots N$ as well as $\bar{x}_j$ and $\bar{p}_j$,
respectively. They are ordered in vectors $\mathbf{x}$,
$\bar{\mathbf{x}}$, $\mathbf{p}$ and $\bar{\mathbf{p}}$. The chains
are coupled by an interaction $H_{\rm{int}}$.  When the coupling is
``switched off" i.e., $H_{\rm{int}}=0$ these two chains are governed
by the Hamiltonians
\begin{equation}
H_{\rm{I}}  = \frac{1}{2m}\left(\mathbf{p},\mathbf{p}\right) 
                   + \left(\mathbf{x}, \potential\ \mathbf{x}\right),\qquad
H_{\rm{II}} = \frac{1}{2m}\left(\bar{\mathbf{p}},\bar{\mathbf{p}}\right) 
                   + \left(\bar{\mathbf{x}}, \potential\ \bar{\mathbf{x}}\right), 
\label{Huncoupled}
\end{equation}
where the notation $\left(\cdot,\cdot\right)$ stands for the scalar
product. In the coordinate representation, we have
\begin{equation}
\left(\mathbf{p},\mathbf{p}\right) = 
\sum_{i=1}^{N} p_i^2,\qquad \left(\bar{\mathbf{p}},\bar{\mathbf{p}}\right) 
= \sum_{i=1}^{N} \bar{p}_i^2,
\end{equation}
while the potential terms describing the interactions of different
particles within the chains can be written as
\begin{equation} 
\left(\mathbf{x}, \potential\ \mathbf{x}\right) = \sum_{i,j=1}^{N} x_i \potential_{ij} x_j,\qquad
\left(\bar{\mathbf{x}}, \potential\ \bar{\mathbf{x}}\right) = \sum_{i,j=1}^{N} \bar{x}_i \potential_{ij} \bar{x}_j.
\end{equation}
We assume that such interactions are given by a shift invariant matrix
$\potential_{ij} = \potential_{(i+n){\rm{mod}} N\ (j+n){\rm{mod}} N }$
which, in addition, satisfies translational symmetry condition
$\sum^N_{i=1}\potential_{ij}=0$. This implies that for uncoupled
chains the non-interacting degrees of freedom are phonons.

After introducing the coupling between the two chains the total
Hamiltonian of the system becomes
\begin{equation}
H = H_{\rm{I}} + H_{\rm{II}} + H_{\rm{int}}, 
\label{Hamiltonian}
\end{equation}
where the interaction term
\begin{equation}
H_{\rm{int}} = \sum_{i,j=1}^{N} K_{ij} \left(x_i - \bar{x}_j\right)^2
= \sum_{i,j=1}^{N} K_{ij} (x_i^2 + \bar{x}_j^2) - 2 \sum_{i,j=1}^{N} K_{ij} x_i \bar{x}_j 
\label{potential}
\end{equation}
is determined by positive symmetric coupling constants $K_{ij}$.  In
what follows, we assume that $H(x, p, \bar{x}, \bar{p})$ is a
non-negative function. This guarantees that the motion of the whole system remains bounded for all times. We notice that we do not make a similar requirement for $H_{\rm{I}}(\mathbf{p},\mathbf{q})$ and $H_{\rm{II}}(\bar{\mathbf{p}},\bar{\mathbf{q}})$.

\subsection{Mapping onto a Caldeira-Leggett-like model}{\label{collectivecoordinate}}

In what follows we study the dynamics of the collective coordinate $X$
defined as the difference between the center of masses of two chains
scaled with the factor $\sqrt{N/2}$
\begin{equation}
X=\frac{1}{\sqrt{2N}}\sum_{i=1}^N x_i - \frac{1}{\sqrt{2N}}\sum_{i=1}^N \bar{x}_i.
\end{equation}
To this end we map the general problem (\ref{Hamiltonian}) of two
coupled chains of harmonic oscillators to a model of Caldeira-Leggett
type, where $X$ is coupled to the ``bath" provided by the remaining
degrees of freedom.  This formulation provides an intuitive
description for the dynamics of the collective coordinate in the
process of transferring energy from $X$ to the bath coordinates.  Such
an interpretation, however, has to be used carefully because the
energy transfer happens inside the full system and a precise
definition of the bath depends not only on the form of the Hamiltonian
(\ref{Hamiltonian}), but also on the choice of the collective
coordinate.

As a first step, we introduce the new set of canonical coordinates  and momenta
\begin{eqnarray}
 c_i = \sum_{n=1}^N A_{in}x_n, \quad\chi_i = \sum_{n=1}^N A_{in}p_n, \qquad\\
 \bar{c}_i = \sum_{n=1}^N A_{in}\bar{x}_n, \quad  \bar{\chi}_i = \sum_{n=1}^N A_{in}\bar{p}_n,
\end{eqnarray}
such that  $H_{\rm{I}}$ and $H_{\rm{II}}$ become diagonal
\begin{equation}
H_{\rm{I}} =
\sum_{i=1}^N \left(\frac{{\chi}_i^2}{2 m} + \frac{m \omega_i^2 {c}^2_i}{2}\right) \ ,\qquad
H_{\rm{II}} =
\sum_{i=1}^N \left(\frac{\bar{\chi}_i^2}{2 m} + \frac{m \omega_i^2 \bar{c}^2_i}{2}\right) \ ,
\end{equation}
where $A_{in}$ are the elements of the matrix $A$ that diagonalizes $\potential$  
\begin{equation}
A\potential\! A^T = \frac{m}{2}\ \Omega^2,\qquad
\Omega = \mathrm{diag}(\omega_1,\dots,\omega_N).
\end{equation}
For the translational invariant matrix $\potential$ used in this model
the diagonalization matrix $A$ is given by \cite{Scheck}
\begin{eqnarray}\label{eigenvectors}
 A_{j1}=\sqrt{\frac{1}{N}},\
A_{jm}=\sqrt{\frac{2}{N}}\cos\left(\frac{\pi (m-1)}{N}\left(j-\frac{1}{2}\right)\right), 
\end{eqnarray}
with indices $m=2,\dots N$ and $j=1,\dots N$.
We now express the interaction part of the two chains in the new
coordinates.  For the first term in equation~(\ref{potential}) we obtain
\begin{equation} 
\sum_{i,j=1}^{N} K_{ij} (x_i^2 + \bar{x}_j^2)=\sum_{n,m=1}^{N} \sum_{i=1}^{N} \hat{k}_i A_{in}A_{im} (c_n c_m + \bar{c}_n \bar{c}_m),
\end{equation}
where we introduced $\hat{k}_i = \sum_{j=1}^N K_{ij}$.  Treating the second
term in equation~(\ref{potential}) in an analogous way we obtain for the
interaction part of the Hamiltonian
\begin{equation} 
H_{\rm{int}}
=
\left(\mathbf{c}, \tilde{K}^\alpha \mathbf{c}\right) + \left(\mathbf{\bar{c}}, \tilde{K}^\alpha \mathbf{\bar{c}}\right) -
 \left(\mathbf{c}, \tilde{K}^\beta \mathbf{\bar{c}}\right) - \left(\mathbf{c}, \tilde{K}^\beta \mathbf{\bar{c}}\right),
\end{equation}
where $\mathbf{c}, \mathbf{\bar{c}}$ are vectors with components $c_n
$, $\bar{c}_n$ and $\tilde{K}^\alpha, \tilde{K}^{\beta}$ are the
matrices defined by
\begin{eqnarray} 
\tilde{K}^\alpha
=
A^T \hat{K} A,\qquad
\tilde{K}^{\beta} = A^T K A,
\end{eqnarray}%
with $\hat{K}_{ij}=\delta_{ij}\hat{k}_j$.
Furthermore, after transforming the coordinates and momenta according to
\begin{eqnarray}
  d_n = \frac{\left(c_n -  \bar{c}_n\right)}{\sqrt{2}},\quad \bar{d}_n 
               = \frac{\left(c_n + \bar{c}_n\right)}{ \sqrt{2}}, \qquad\\ \eta_n 
               = \frac{\left(\chi_n -  \bar{\chi}_n\right)}{ \sqrt{2}}, \quad\bar{\eta}_n 
               = \frac{\left(\chi_n +  \bar{\chi}_n\right)}{ \sqrt{2}}
\end{eqnarray}
and defining $\tilde{K}= \tilde{K}^\alpha + \tilde{K}^{\beta}$,
$\bar{K} = \tilde{K}^\alpha - \tilde{K}^{\beta}$, the interaction term
can be cast into the form
\begin{eqnarray} 
H_{\rm{int}}
=
\big(\mathbf{d},\tilde{K}\mathbf{d}\big) + \big(\mathbf{\bar{d}},  \bar{K} \mathbf{\bar{d}}\big).
\end{eqnarray}
With this new set of canonical coordinates the Hamiltonian becomes
\begin{eqnarray} 
 H 
&=&
\sum_{i=1}^N \left(\frac{\eta^2_i}{2 m} + \frac{\bar{\eta}^2_i}{2 m}\right) 
     + \frac{m}{2} \sum_{i=1}^N \left(\omega^2_i \bar{d}^2_i + \omega^2_i d^2_i\right)\nonumber\\
 & &\qquad\qquad + \sum_{n,m=1}^N \tilde{K}_{nm} d_n d_m + \sum_{n,m=1}^N \bar{K}_{nm} \bar{d}_n \bar{d}_m 
\nonumber\\
 &=&
\frac{1}{2m}  \left(\mathbf{\eta}, \mathbf{\eta}\right)  + \left(\mathbf{d}, \frac{m}{2}\Omega^2 + \tilde{K}\mathbf{d}\right)\nonumber\\ 
& &\qquad\qquad + \frac{1}{2m}  
 \left(\bar{\mathbf{\eta}} ,\bar{\mathbf{\eta}}\right)
  + \left(\bar{\mathbf{d}}, \frac{m}{2}\Omega^2 +\bar{K} \bar{\mathbf{d}}\right)   \label{hamilt16}.
\end{eqnarray}
We notice that the collective coordinate and momentum are just
\begin{eqnarray} 
X=  d_1, \qquad P= \eta_1
\end{eqnarray}
and the corresponding frequency is $\omega_1=0$.  Since $X$ couples
only to the coordinates $d_n$, the part of $H$ which depends on
$\bar{\mathbf{d}}, \bar{\mathbf{\eta}}$ can be disregarded when the
dynamics of the collective mode is considered.  Consequently, the
relevant part of the Hamiltonian is given by
\begin{eqnarray}{\label{hamiltonian}}
 H'
&=&
\frac{P^2}{2 m} + {\tilde{K}_{11} X^2} + {X} \sum_{n=2}^N \tilde{K}_{n1} d_n +  \sum_{n = 2}^N \frac{\eta^2_n }{2m}\nonumber\\
& & \qquad\qquad\qquad + \sum_{m,n = 2}^N  \left(\tilde{K}_{nm} + \frac{m \omega^2_n \delta_{nm}}{2}\right) d_n d_m. \label{hamilt18}
\end{eqnarray} 
This already strongly resembles the Caldeira-Legget model but with the
non-diagonal bath Hamiltonian
\begin{eqnarray}
H_{\rm{bath}} 
&=&
\sum_{n = 1}^{N-1} \frac{\eta^2_{n+1}}{2m} + \sum_{n,m = 1}^{N-1}B_{nm} d_{n+1} d_{m+1}, 
\end{eqnarray}
where the elements of the $(N-1)\times (N-1)$ matrix $B$ are given by 
\begin{equation} 
B_{(n-1)(m-1)}=\tilde{K}_{nm} + \frac{m \omega^2_{n} \delta_{nm}}{2}, \qquad n,m=2,\dots N.
\end{equation}
To cast the bath Hamiltonian into diagonal form, we introduce yet
another set of the coordinates $\xi_i = \sum_{j=1}^{N-1}U_{ji}
d_{j+1}$, $\nu_i = \sum_{j=1}^{N-1}U_{ji} \eta_{j+1}$, where $U$ is
the orthogonal matrix diagonalizing $B$,
 \begin{equation}
U^T B U = \frac{m}{2}\ \tilde{\Omega}^2, \qquad \tilde\Omega=\mbox{diag}(\tilde{\omega}_1,\dots\tilde{\omega}_{N-1}).
\end{equation}
With this choice of  coordinates $H_{\rm{bath}}$ reads
\begin{eqnarray}
H_{\rm{bath}}
&=&
\sum_{n=1}^{N-1} \left(\frac{\nu_n^2}{2m} + \frac{m}{2}\ \tilde{\omega}_n^2 \xi_n^2\right),
\end{eqnarray}
Now we have to perform the transformation in the part of the
Hamiltonian that represents the interaction between the bath
coordinates and the collective degree of freedom,
\begin{eqnarray}
{X} \sum_{n=1}^{N-1} {k}_{n} d_{n+1} 
=
{X} \sum_{n,m=1}^{N-1} U_{nm} {k}_{n} \xi_m
=
{X} (\mathbf{l}, \mathbf{\xi}) \, ,
\end{eqnarray}
where we defined the vectors $\mathbf{k}$ and $\mathbf{l} = U^T
\mathbf{k}$ with the components
\begin{eqnarray}
l_m =  \sum_{n=1}^{N-1} {U}_{mn}^T k_n, \qquad k_n = \tilde{K}_{1 (n+1)}, \quad n=1,\dots N-1. 
\end{eqnarray}
Putting all the expressions together we finally arrive at the
following Caldeira-Legget form for our model
\begin{eqnarray}
H' = \frac{P^2}{2m} + {\tilde{K}_{11}} X^2 
             + {X} \sum_{m=1}^{N-1} {l}_m \xi_m 
             + \sum_{n=1}^{N-1} \left(\frac{\nu_n^2}{2m} 
             + \frac{1}{2} { m \tilde{\omega}_n^2 \xi_n^2}\right)  . 
\label{thehamiltonian}
\end{eqnarray}
This Hamiltonian describes an effective particle moving in a harmonic
potential and also interacting with a heat bath. We emphasize again
that contrary to the Caldeira-Legget model the bath is part of the
system and not an external configuration of particles.  The damping of
the collective motion is a result of a redistribution of energy and
not an actual loss of energy as in models with an external bath.
Furthermore, our model possesses only a finite number of degrees of
freedom which eventually causes a return of energy into the collective
mode. This recurrence time will be much longer, however, than the
spreading time for sufficiently large number of particles.

\section{Chain of Oscillators with Next Neighbor Coupling as an
 Example }{\label{Sec3}}

Below  we illustrate the above mapping
procedure for a simple example, where the resulting Hamiltonian
(\ref{thehamiltonian}) can be written down explicitly.  We consider a
system of two chains with next neighbor interaction coupled at one point.
The Hamiltonian for that system reads
\begin{eqnarray}
H &=& \sum_{j=1}^N \frac{1}{2m}  (p_j^2 + \bar{p}_j^2) + \frac{\alpha}{2}
(x_1-\bar{x}_1)^2  \nonumber\\
   & & \qquad +\frac{m \omega^2_0}{2} \sum_{j=1}^{N}
          \Bigl((x_j-x_{(j+1)\mathrm{mod} N})^2 +
(\bar{x}_j-\bar{x}_{(j+1)\mathrm{mod} N})^2\Bigr)
                            \ , \label{example}
\end{eqnarray}
where $K_{ij} =  ({\alpha}/{2})\delta_{i1}\delta_{j1}$ are the coupling
constants. The eigenfrequencies for  a  free chain
of $N$ oscillators with the next-neighbor interaction  as in
(\ref{example})  are given by  \cite{Scheck}
\begin{eqnarray}
\omega_k = 2 \omega_0 \left|\sin\frac{\pi (k-1)}{2N}\right| \ ,  \quad
k=1,...,N.
\end{eqnarray}
with the corresponding eigenvectors  given by (\ref{eigenvectors}).  As
described in section
\ref{collectivecoordinate} we define the set of new coordinates $d_i$  and
 consider the part of the Hamiltonian $H'$ which only
contains the couplings between the $d_j$'s and $X=d_1$.  Straightforward
calculations  then yield
\begin{eqnarray}
H' = \frac{P^2}{2m} + \frac{\alpha}{N}X^2
         + \frac{\alpha}{\sqrt{N}}X (\mathbf{a}, \mathbf{d}) +
H_{\rm{bath}}\ ,
\end{eqnarray}
where $\mathbf{a} = (A_{1 2},...,A_{1 N})$ and
the bath Hamiltonian is given by
\begin{eqnarray}
H_{\rm{bath}} =  \frac{1}{2m}(\mathbf{\eta}, \mathbf{\eta})
 + \left(\mathbf{d}, \left(  \frac{m}{2}{\Omega}^2
 + \alpha\ \mathbf{a}\otimes\mathbf{a}\right) \mathbf{d}\right)
\end{eqnarray}
with
${\Omega}^2$ being the diagonal matrix of the eigenvalues $\omega_k^2$
and ``$\otimes$'' being the ordinary tensor product.  Diagonalization
of the bath leads to
\begin{eqnarray}
H' = \frac{P^2}{2m} + \frac{2\alpha}{N} X^2 +  X \sum_{n=2}^N C_n(\alpha)
\xi_n
+ \sum_{n=2}^N\left( \frac{\nu_n^2}{2m} + \frac{m \tilde{\omega}_n^2
\xi_n^2}{2} \right)
\end{eqnarray}
with the coupling coefficients
\begin{eqnarray}
C_n(\alpha) =
\frac{\sqrt{2}\alpha}{N}   \left(\sum_{k =2}^N
\frac{ \cos^2(\frac{\pi (k-1)}{2N}) }{(\tilde{\omega}^2_n -
\omega^2_k)^2}\right)^{-{1}/{2}}
\sum_{k =2}^N\frac{{\cos^2(\frac{\pi (k-1)}{2N})}}{\tilde{\omega}^2_n -
\omega^2_k} \ ,
\end{eqnarray}
where the implicit equation 
\begin{eqnarray}
\frac{4\alpha}{Nm} \sum_{k=2}^N
\frac{\cos^2\left(\frac{\pi(k-1)}{2N}\right)}{\tilde{\omega}^2_j -
\omega_k^2} = 1 
\end{eqnarray}
yields the eigenfrequencies $\tilde{\omega}_j$.

\section{ Dynamics of the Collective Coordinate  }{\label{Sec4}}

We return to the general case.  So far we mapped the Hamiltonian
system of two coupled chains of harmonic oscillators to the
Caldeira-Leggett model.  The next step is to consider the time
evolution of the collective mode $X(t)$ induced by the Hamiltonian
(\ref{thehamiltonian}). The full quantum mechanical solution of the
problem would require calculating the time evolution for a reduced
density-matrix $\hat{\rho}_{\mathbf{\scriptscriptstyle{rd}}}(X)$ of
the collective coordinate. While such an analysis is certainly
possible along the lines of Ref.~\cite{Caldeira:1982iu, Aurel2}, for our
purposes it will be sufficient to consider the most basic collective
dynamical properties captured by the time evolution of the expectation
value for the quantized collective observable $\hat{X}$
\begin{eqnarray}
\langle{\hat{X}(t)}\rangle :=\mbox{Tr}(\hat{\rho} \hat{X}(t)) \ ,
\label{expectX}
\end{eqnarray}
where $\hat{\rho}$ is the full density matrix.  In this case the
problem simplifies, since one can deduce the time evolution equation
for $\langle\hat{X}(t)\rangle$ from the corresponding equation for the
time evolution of the quantum operator $\hat{X}(t)$ \cite{Breuer}. It
is worthwhile to mention that, since $\hat{H}'$ contains only
quadratic terms, the resulting equation of motion for
$\langle{\hat{X}(t)}\rangle$ coincides with the corresponding equation
of motions for the classical observable $X(t)$ obtained for the
classical Hamiltonian $H'$. Below we give a short derivation of this
equation and analyze its solution for certain types of initial
conditions for $\hat{\rho}$.

The Heisenberg equations for our system read
\begin{eqnarray}
\dot{\hat{X}}(t) = \frac{i}{\hbar}[\hat{H}',\hat{X}(t)] = \frac{\hat{P}}{m},   
\end{eqnarray}
\begin{eqnarray}
\dot{\hat{P}}(t) = \frac{i}{\hbar}[\hat{H}',\hat{P}(t)] = -2\tilde{K}_{11} \hat{X} + \sum_{n=1}^{N-1} l_n \hat{\xi}_n(t),  
\end{eqnarray}
\begin{eqnarray}
\dot{\hat{\xi}}_n(t) = \frac{i}{\hbar}[\hat{H}',\hat{\xi}_n(t)] = \frac{\hat{\nu}_n}{m},  
\end{eqnarray}
\begin{eqnarray}
\dot{\hat{\nu}}_n(t) = \frac{i}{\hbar}[\hat{H}',\hat{\xi}_n(t)] = - m \tilde{\omega}_n^2 \hat{\xi}_n(t) + l_n\hat{X}(t).
\end{eqnarray}
From these equations one immediately  obtains 
\begin{equation} 
m\ddot{\hat{X}}(t) + 2 \tilde{K}_{11} \hat{X}- \sum_{n=1}^{N-1} l_n \hat{\xi}_n(t) = 0
\end{equation}\
and
\begin{equation}{\label{a}} 
 m\ddot{\hat{\xi}}_n(t) + m \tilde{\omega}_n^2 \hat{\xi}_n(t) - l_n \hat{X}(t) = 0, \qquad n=1,\dots N-1.
\end{equation}
We now use the representation of the momentum and coordinate operators
at time zero in terms of creation and annihilation operators
\begin{equation}{\label{b}}  
\hat{\xi}_n(0) = \sqrt{\frac{\hbar}{2 m \tilde{\omega}_n}} (b_n + b_n^\dagger), \qquad
\hat{\nu}_n(0) =-i \sqrt{\frac{m \hbar \tilde{\omega}_n}{2}} (b_n - b_n^\dagger).
\end{equation}
With these initial conditions the  solution of equation (\ref{a}) takes the form
\begin{eqnarray}
\hat{\xi}_n(t) &=& 
\sqrt{\frac{\hbar}{2 m \tilde{\omega}_n}} (e^{-i\tilde{\omega}_n t} b_n 
  + e^{i\tilde{\omega}_n t} b_n^\dagger)\nonumber\\ 
& &\qquad\qquad\qquad  + \frac{l_n}{m \tilde{\omega}_n} \int\limits_{0}^{t} ds\ \sin(\tilde{\omega}_n(t-s)) \hat{X}(s).
\end{eqnarray}
Using  this  to eliminate the bath-modes from the equation (\ref{b}), we obtain
\begin{eqnarray}
\ddot{\hat{X}}(t) + \frac{2\tilde{K}_{11}}{m}  \hat{X}
       - \frac{2}{m} \int\limits_{0}^{t}\int\limits_{0}^{\infty}ds\ d\tilde{\omega}\   
           \sigma(\tilde{\omega}) \sin(\tilde{\omega}(t-s))\  \hat{X}(s)
= \frac{ \hat{F}(t)}{m} 
\label{intermeq}
\end{eqnarray}
where 
\begin{eqnarray}
 \hat{F}(t) =
\sum_{n=1}^{N-1} l_n \sqrt{\frac{\hbar}{2m \tilde{\omega}_n}} 
         (e^{-i\tilde{\omega}_n t} b_n + e^{i\tilde{\omega}_n t} b_n^\dagger)
\end{eqnarray}
is the force operator that acts on the collective coordinate and
\begin{eqnarray}{\label{density}}
\sigma(\tilde{\omega}) =
\sum_{n=1}^{N-1} \frac{l_n^2}{2 m \tilde{\omega}_n}\ \delta(\tilde{\omega} - \tilde{\omega}_n)
\end{eqnarray}
is the spectral density.  We further rewrite the part describing the
dissipation as
\begin{eqnarray}
- \frac{2}{m} \int\limits_{0}^{t}\int\limits_{0}^{\infty}ds\ d\tilde{\omega}\   \sigma(\tilde{\omega})  \ \sin(\tilde{\omega}(t-s))  \hat{X}(s) =
\int\limits_{0}^{t} \frac{d}{dt}\gamma(t-s)\  \hat{X}(s) ds,
\end{eqnarray}
where we defined the damping-kernel as
\begin{eqnarray}
\gamma(t-s) 
=
\frac{2}{m}\int\limits_{0}^{\infty} d\tilde{\omega}\ \frac{\sigma(\tilde{\omega})}{\tilde{\omega}}\ \cos(\tilde{\omega}(t-s)).
\end{eqnarray}
After inserting this term into equation~(\ref{intermeq}) we arrive at
\begin{eqnarray}
\frac{d^2\hat{X} (t)}{dt^2}   + \frac{2\tilde{K}_{11}}{m}\hat{X}(t)  
     + \int\limits_0^t ds\ \dot{\gamma}(t-s) \hat{X}(s)= 
\frac{1}{m} \hat{F}(t) \ . 
\label{qevolution}
\end{eqnarray}
We now use equation (\ref{qevolution}) to obtain the evolution
equation for the expectation value (\ref{expectX}) of $X$ for some
class of initial states $\hat{\rho}$.  We assume that the initial
conditions for $\hat{\rho}$ satisfies
\begin{equation}
\langle{\hat{X}(0)}\rangle =0 \ , \qquad \langle{\hat{P}(0)}\rangle=P_0 \ , \qquad 
\langle b_n\rangle=\langle b_n^\dagger\rangle=0 \ . 
\label{initialcond}
\end{equation}
Here we have used the notation $\langle
\hat{A}\rangle:=\mbox{Tr}(\hat{\rho} \hat{A})$ for the expectation
value of an observable $ \hat{A}$.  Under these assumptions
equation~(\ref{qevolution}) yields for the expectation value of
$\hat{X}$
\begin{eqnarray}{\label{dynamics}}
\frac{d^2\langle\hat{X}(t)\rangle  }{dt^2}  + \Omega_0^2 \langle\hat{X}(t)\rangle    + \int\limits_0^t ds\ \gamma(t-s) \frac{d\langle{\hat{X}}(s)\rangle }{ds} 
= 0 \ ,
\label{expectationeq}
\end{eqnarray}
where $\Omega_0^2 = {2\tilde{K}_{11}}/{m}- \gamma(0)$ and the term
$\gamma(0)$ is a renormalization of the potential resulting from the
interaction between the collective mode and the bath. Equation
(\ref{expectationeq}) is a classical damping equation which together
with the initial conditions (\ref{initialcond}) describes the time
development of the collective mode.  It is straightforward to see that
one obtains precisely the same equation for classical time evolution
of $X$ under the classical Hamiltonian flow induced by $H'$ if the
initial conditions are fixed as
\begin{equation}
X(0)=0 \ , \quad P(0)=P_0 \ , \quad \xi_i=0 \ , \quad \nu_i=0 \ , \quad i=1,\dots N-1 \ .
\end{equation}
We notice that the entire information on the time evolution of
$\langle\hat{X}(t)\rangle$ is encoded in the damping kernel $\gamma$.
If $\gamma(t)=\gamma_0\delta(t)$, that is, if the system has no
``memory'', the above equation describes the damped harmonic
oscillator of frequency $\Omega_0$ with the damping coefficient
$\gamma_0$.

Since (\ref{expectationeq}) is a linear equation, we can easily
construct its solution for a general kernel $\gamma(t)$. To this end
we consider a slightly different equation
\begin{eqnarray}
\frac{d^2\langle{\hat{X}}(t)\rangle }{dt^2}  + \Omega_0^2 \langle\hat{X}(t)\rangle    
 + \int\limits_{-\infty}^{\infty} ds\  \theta(t-s)\gamma(t-s) \frac{d\langle{\hat{X}}(s)\rangle }{ds}  
= \frac{P_0}{m}\ \delta(t), \label{modifiedeq}
\end{eqnarray}
with  the initial conditions  
\begin{eqnarray}{\label{initialconditions}}
\langle \hat{X}(-\infty)\rangle = 0, \qquad \langle \hat{P}(-\infty)\rangle = 0.
\end{eqnarray}
at time $t=-\infty$.  Equation (\ref{modifiedeq}) describes thus the
system which stays at rest for all times $t<0$ and then gets a ``kick"
at the time $t=0$. After this it acquires a momentum $P_0$ and
continues to evolve according to equation~(\ref{expectationeq}).
Obviously both, equation~(\ref{expectationeq}) and
equation~(\ref{modifiedeq}), give the same solution for positive
times. We can solve equation~(\ref{modifiedeq}) employing the pair of Fourier
transforms
\begin{eqnarray}
\langle\hat{X}(t)\rangle =
\int\limits_{-\infty}^{\infty} \tilde{X}(\omega)\ e^{-i\omega t} d\omega,\qquad
\tilde{X}(\omega) =
\frac{1}{ 2\pi} \int\limits_{-\infty}^{\infty} \langle\hat{X}(t)\rangle\ e^{i\omega t} dt.
\end{eqnarray}
Applying the Fourier transformation to both sides of
equation~(\ref{modifiedeq}) we find the following expression
\begin{eqnarray}\label{FT1}
\tilde{X}(\omega) =
\frac{P_0}{2\pi m (\Omega_0^2 - \omega^2 -  i \omega\tilde{\gamma}(\omega))},
\end{eqnarray}
where $\tilde{\gamma}(\omega)$ is defined as
\begin{eqnarray}
\tilde{\gamma}(\omega) :=
\int\limits_{0}^{\infty}  \gamma(s) e^{i\omega s} ds. \label{gamma2omega}
\end{eqnarray}
Therefore the solution of the homogeneous system becomes
\begin{eqnarray}
\langle\hat{X}(t)\rangle =
\frac{P_0}{2\pi m}\int\limits_{-\infty}^{\infty}
            \frac{\ e^{i\omega t}}{\Omega_0^2 - \omega^2 -  
            i \omega \tilde{\gamma}(\omega)}\ d\omega \ . 
\label{solution}
\end{eqnarray}
As one can see from equations~(\ref{gamma2omega}) and
(\ref{solution}), the dynamics of the collective mode is encoded in
the spectral density $\sigma(\omega)$. It is thus important to relate
$\sigma(\omega)$ to the interaction matrix $\tilde{K}$ appearing in
the original Hamiltonian (\ref{hamiltonian}). Recalling the definition
(\ref{density}) of $\sigma$ and using $\mathbf{k} = U \mathbf{l}$ we
obtain
\begin{eqnarray}
 \sigma(\omega)
&=&
-\frac{1}{2\pi m \omega} \mathrm{Im} \left(\sum_{n=1}^{N-1}  
       \frac{{l}_n {l}_n^*}{\omega-\tilde{\omega}_n + i \epsilon}\right)\nonumber\\  
 &=&
-\frac{1}{2\pi m \omega} \mathrm{Im}\left( \sum_{n=1}^{N-1} 
       \frac{ \left[{\mathbf{l} \otimes \mathbf{l}^T}\right]_{n,n}}{\omega-\tilde{\omega}_n 
      + i \epsilon} \right)\nonumber\\
& =&
-\frac{1}{2\pi m\omega}  \mathrm{Im} \mathrm{Tr}
        \left[ \frac{ {\mathbf{k} \otimes \mathbf{k}^T}}{\omega\eins- 
           \left(\frac{2}{m}B\right)^{{1}/{2}} + i \epsilon}\right] \ ,
\end{eqnarray}
where $\mathbf{l} \otimes \mathbf{l}^T$, $\mathbf{k} \otimes
\mathbf{k}^T$ stands for the tensor product between $\mathbf{l}$ and
$\mathbf{l}^T$ (resp. $\mathbf{k}$ and $\mathbf{k}^T$).  The last
expression can be rewritten in terms of a scalar product,
\begin{eqnarray}
\sigma(\omega) =
-\frac{1}{2\pi m\omega} \mathrm{Im}\left( \mathbf{k}, \,
         \frac{1}{\omega\eins- ({\Omega_r^2} + \frac{2}{m}\tilde{K}_r)^{1/2} 
        + i \epsilon} \,\, \mathbf{k}\right) \ , 
\label{densityexpression}
\end{eqnarray}
where $\Omega_r$, $\tilde{K}_r$ are $(N-1)\times(N-1)$ matricies
obtained from $\Omega$, $\tilde{K}$ by deleting the first row and the
first column, respectively.  We have now a formal expression for the
spectral density of our general model.  Two remarks are in order.
First the collective coordinate becomes completely decoupled from the
bath if and only if $\mathbf{k}=0$. Since the components of
$\mathbf{k}$ can be written as
\begin{equation}
k_i=\frac{2}{\sqrt{N}}\sum_{j=1}^N \hat{k}_jA_{j(i+1)} \ ,
\end{equation}
the above condition is equivalent to the requirement that the
$\hat{k}_i=\sum_{j=1}^{N} K_{ij}$ take the same value for all $i$. In
particular, there is no damping if $K_{ij}=const$.  We notice that
given a splitting of the interactions: $K_{ij}=\mathcal{K}+\delta
K_{ij}$ into ``constant" and ``fluctuating" parts of the interaction,
only $\delta K_{ij}$ contributes to $\mathbf{k}$.  Second, by adding
the term $K_0 X^2$ to the Hamiltonian (\ref{Hamiltonian}) one can
adjust the collective frequency $\tilde{\Omega}_0$ without changing
the spectral density $\sigma$. This additional term can be incorporated
into $H_I, H_{II}$, $H_{\mathrm{int}}$ such that the overall structural
form of $H$ remains intact. Note that this ``renormalization" results
in a shift of the spectrum $ \Omega_r$ of the chain Hamiltonians $H_I,
H_{II}$ which is compensated by the shift of the interaction term
$\tilde{K}_r$ by a diagonal matrix, such that the matrix $B$ (resp.
$\sigma$) does not change.

The form (\ref{densityexpression}) for the density $\sigma$ hinders an
exact treatment for a general form of interaction matrix $K$. However,
if we assume that the fluctuation part of couplings matrix elements
are small $|\delta K_{ij}| \ll m|\omega^2_{n+1}-\omega^2_{n}|$, we
can approximate the density function by
\begin{equation}
  \sigma(\omega) = \sum_{n=1}^{N-1} \frac{k_n^2}{2 m \omega}\ 
  \delta\left(\omega - \sqrt{{\omega}^2_n+{2N}\mathcal{K}/m}\right) \ ,
\label{sigexp}
\end{equation}
where $\{\omega_n\}$ is the phononic spectrum of the noninteracting
chains and the $k_n$'s are determined solely by $\delta K_{ij}$. The
expression (\ref{sigexp}) can be interpreted to the extent that after
introducing the interaction between the two chains the phonons acquire
a ``mass". Assuming that $k_n$ are uniformly distributed, the behavior
of $\sigma(\omega)$ is determined by the spectral density of the
phonon frequencies $\omega_n$. In particular, in the case of an Ohmic law distribution for the $\omega_n$ this leads to $\sigma(\omega)\sim\omega\Theta(\omega-\frac{2N}{m}\mathcal{K})$ at
low frequencies.  Furthermore, if $\mathcal{K}=0$ this in turn implies
that $\gamma(t)$ is localized at $t=0$ and
equation~(\ref{expectationeq}) can be approximated by the differential
equation describing time evolution of a harmonic oscillator with a
friction.

\section{Transition strengths and collective excitation}{\label{Sec6}}

In the previous section, we derived an equation of motion that describes
the damping of the collective excitation. As we mentioned already, the
quantum evolution governed by equation~(\ref{expectationeq}) coincides
with the classical evolution of $X(t)$ if the initial conditions are
defined in an appropriate way. In this section we consider the problem
of existence of quantum collective states in the spectrum of the
system.  One way to probe such collective excitations is to couple the
system to an external weak periodic potential $v(X,t)\sim A(
X)\cos(\omega t)$ depending on the collective variable $X$. Assuming
that the coupling is weak, the energy absorption rate in the first
order perturbation theory will be determined by the following spectral
function
\begin{equation}
 \tilde{S}_A(\omega)=\sum_{n=1}^N |\vacl A(\hat{X}) |n\rangle|^2\,
         \delta\left(\omega-\frac{E_n -E_0}{\hbar}\right),\label{specfunc}
\end{equation}
with $T_n=|\vacl A(\hat{X}) |n\rangle|^2$ being the transition
strengths between the ground state with energy $E_0$ and $n$-th state
with energy $E_n$. The collective states can then be defined, as
states having large transition strengths $T_n$.  Accordingly, the
spectral function (\ref{specfunc}) keeps the information about the
existence of collective modes in the system.  Equivalently, one can
consider the Fourier transform of $\tilde{S}_A(\omega)$, which is
given by the time correlation of $A(\hat{X})$
\begin{equation}
 S_A(t)= \vacl A(\hat{X}(t))A(\hat{X}(0))\vacr.
\end{equation}
On an intuitive level one might expect that the averaged transition
strengths $T_n$ should exhibit spikes for the energies $E_n$
corresponding to collective motion.  Below we show that under certain
conditions this is indeed the case and the dynamical equation
(\ref{expectationeq}), in fact, determines the form of the time
correlations $S_A(t)$.

\subsection{Transition strengths induced by $\hat{X}$}{\label{Sec61}}

Let us first consider the case of the observable $A(X)=X$. We
calculate the time correlator
\begin{equation}
 S(t)= \vacl \hat{X}(t) \hat{X}(0)\vacr. \label{Xcorrelator}
\end{equation}
Since we are dealing here with a system of coupled harmonic
oscillators it is useful to consider the set of normal coordinates
$(q_n, p_n)$ where the Hamiltonian (\ref{hamilt16}) becomes diagonal \cite{Feynman},
\begin{eqnarray}
\hat{H} =
\sum_{i=1}^{2N} \left({\frac{\hat{p}_i^2}{2m}} + \frac{m\omeg_i^2 \hat{q}_i^2}{2}\right)
= \sum_{i=1}^{2N} \hbar \omeg_i \left(\hat{a}^\dag_i \hat{a}_i + \frac{1}{2}\right) 
\end{eqnarray}
Here $\hat{q}_i$, $\hat{p}_i$ are the position and the momentum
operators corresponding to $(q_i, p_i)$, with $\hat{a}^\dag_i$,
$\hat{a}_i$ being the creation and the annihilation operators,
respectively. The frequencies $\omeg_i$ are the eigenfrequencies of
the full system. Since the connection between old coordinates $X$,
$\{d_i\}$, $\{\bar{d}_i\}$ and new $\{q_i\}$ coordinates is given by a
linear transformation, we can assume that
\begin{eqnarray}{\label{hallo}}
\hat{X} = \sum_{i=1}^{2N}  \const_i \hat{q}_i
\end{eqnarray}
with some coefficients $\const_i$.  Substituting (\ref{hallo}) into
(\ref{Xcorrelator}) we obtain
\begin{eqnarray} 
S(t) &=& \vacl\hat{X}(t) \hat{X}(0) \vacr \nonumber\\
        &=& \vacl \exp(i\hat{H}t/\hbar) \hat{X}(0) 
          \exp(-i\hat{H}t/\hbar) \hat{X}(0) \vacr \nonumber\\ 
        &=& \sum_{n=1}^{2N}  |\vacl\hat{X}(0)|n\rangle|^2 \exp\left(i\frac{(E_0-E_n)t}{\hbar}\right)\nonumber\\
        &=& \frac{\hbar}{2m}\sum_{n=1}^{2N} \frac{\const_n^2}{\omeg_n} \exp(-i\omeg_n t) \ ,
\end{eqnarray}
where we used the relations $\hat{q}_i =
\sqrt{\hbar/2m\omeg_i}(\hat{a}^\dag_i + \hat{a}_i)$ to calculate the
transition strength between the ground state $|0_1 0_2\ldots 0_{2N}
\rangle =\vacr$ and excited states $|n_1 n_2\ldots n_{2N} \rangle =
|n\rangle$. Taking then the Fourier transform of $S(t)$ leads to
\begin{eqnarray}{\label{FT}}
\tilde{S}(\omega) =
\sum_{n=1}^{2N}  |\vacl\hat{X}(0)|n\rangle|^2  \delta(\omega - \omeg_n)
= \frac{\hbar}{2m}\sum_{n=1}^{2N} \frac{\const_n^2}{\omeg_n}  \delta(\omega - \omeg_n) \ .
\label{eqforSt}
\end{eqnarray}
Although $\tilde{S}(\omega)$ is a quantum mechanical object, we will
show now that it is possible to relate it to the dynamics of a purely
classical damped harmonic oscillator. To this end we consider the time
evolution of the collective coordinate $X$ under the Hamiltonian $H$
with the following initial conditions:
\begin{eqnarray}{\label{incond}}
\dot{X}(0) = \frac{P_0}{m}, \quad  X(0) = 0, \qquad d_i=0, \,\,\dot{d_i}(0)=0, \,\, \forall i> 1. 
\end{eqnarray}
As has been explained in the previous section, the dynamical evolution
of $X(t)$ with such boundary conditions is governed by
equation~(\ref{expectationeq}) for the classical damped oscillator. On
the other hand, we can express this solution in the diagonalizing
coordinates $q$ as follows.  The time evolution of $q_n(t)$ is given
by
\begin{eqnarray}
q_n  = A_n \sin(\omeg_n t) \ .
\end{eqnarray} 
where the constants $A_n$ are fixed by  the initial conditions (\ref{incond}): 
\begin{eqnarray}
\dot{q}_n(0) = A_n \omeg_n = \frac{P_0}{m} \const_n^*.
\end{eqnarray} 
Accordingly, for the time evolution of $X(t)$ we  obtain
\begin{eqnarray}
X(t) = \sum_{n=1}^{2N} \const_n q_n(t)= 
\frac{P_0}{m}\sum_{n=1}^{2N} \frac{|\const_n|^2}{\omeg_n} \sin(\omega_n t). \label{eqforX}
\end{eqnarray} 
Comparing equations~(\ref{eqforX}) and~(\ref{eqforSt}), we see that
the classical quantity $X(t)$ and the imaginary part of $S(t)$ are
related via
\begin{eqnarray}
S_1(t):=\mathrm{Im} S(t) = 
-\frac{\hbar}{2m}\sum_{n=1}^{2N} \frac{|\const_n|^2}{\omeg_n} \sin(\omeg_n t)
= -\frac{\hbar}{2P_0} X(t).
\end{eqnarray}
Taking the Fourier transform of $S_1(t)$ yields 
\begin{eqnarray}
\tilde{S}_1(\omega) =
\frac{i\hbar}{2m} \sum_{n=1}^{2N} \frac{|\const_n|^2}{2\omeg_n}(\delta(\omega-\omeg_n)
        -\delta(\omega +\omeg_n))=
-\frac{i\hbar}{P_0} \mathrm{Im}\tilde{X}(\omega)
\end{eqnarray}
where $\tilde{X}(\omega)$ is given by the righthand side of
equation~(\ref{FT1}).  Furthermore, comparing this expression with
(\ref{FT}) we recognize the connection
\begin{eqnarray}
\tilde{S}(\omega) =
2 i \theta(\omega) \tilde{S}_1(\omega)=
 \frac{2\hbar}{P_0}\theta(\omega) \mathrm{Im}\tilde{X}(\omega) , \label{s1def}
\end{eqnarray}
where $\theta(\omega)$ denotes the Heaviside step function. This can
be also written explicitly as
\begin{eqnarray}
\tilde{S}(\omega)
= \frac{\hbar}{2\pi m}\theta(\omega)\mathrm{Im}\left(\frac{1}{{\Omega}_0^2 - {\omega}^2 -  i {\omega} \tilde{\gamma}({\omega})} \right). \label{tildeSomega}
\end{eqnarray}
It is worth noticing that this expression for $\tilde{S}(\omega)$ can
also be derived using the fluctuation-dissipation theorem. Suppose
at a certain moment a weak time dependent perturbation
$\delta \hat{H} =\hat{X}\f (t)$ is added to the Hamiltonian
(\ref{Hamiltonian}). Under this external perturbation the system will
be driven away from the ground state. Considering the linear response
of the system to $\delta \hat{H}$, it follows (see e.g.,
\cite{Breuer}) that the averaged displacement of the collective
coordinate is given by
\begin{equation}
\langle \hat{X}(t)\rangle =\int\limits_{-\infty}^{\infty} dt'\, \chi(t-t') \f (t'),\label{fluctdis}
\end{equation} 
where the integration kernel is given by
$\hbar\chi(t)=-2\theta(t)\mathrm{Im}\vacl \hat{X}(t)
\hat{X}(0)\vacr= -2\theta(t)S_1 (t)$. On the other hand,
from the previous section we know that for any force $\f (t)$ (not
necessary weak) the evolution of $\langle \hat{X}(t)\rangle$ is
described by the equation:
\begin{eqnarray}
\frac{d^2\langle\hat{X}(t)\rangle  }{dt^2}  
 + \Omega_0^2 \langle\hat{X}(t)\rangle    
  + \int\limits_0^t ds\ \gamma(t-s) \frac{d\langle{\hat{X}}(s)\rangle }{ds} 
= \frac{\f (t)}{m}. 
\end{eqnarray}
Taking the Fourier transform from both sides of this expression and
comparing the result with the Fourier transformed
equation~(\ref{fluctdis}) leads then to (\ref{tildeSomega}).

From equation~(\ref{tildeSomega}) we clearly see that the
information on the distribution of the transition strengths is stored
in the damping kernel $\gamma(t)$ of the purely classical equation for
the time evolution of the collective mode. One should note, however,
that $ \tilde{S}(\omega)$ is not a smooth function but a sum of
distributions with wildly fluctuating strength. It is easy to see, for
instance, that most of the states are actually not coupled at all to
the ground state through the operator $\hat{X}$. Thus, in order to see
a structural emergence of collective excitations, we need to consider
a smoothened version of the spectral function $ \tilde{S}(\omega)$
where the average is taken over some interval
$[\omega-\Delta\omega/2,\omega+\Delta\omega/2]$, such that
$\Delta\omega \gg \delta\omeg$, with
$\delta\omeg:=|\omeg_{n+1}-\omeg_{n}|$ being the difference between
two adjacent frequencies. We can define such a smoothened spectral
function as the convolution
\begin{equation}
\tilde{S}^{(\varepsilon)}_1(\omega):= 
\frac{1}{\pi}\int\limits_{-\infty}^{\infty} d\tilde{\omega}
     \frac{\varepsilon \tilde{S}_1(\tilde{\omega})}{(\omega-\tilde{\omega})^2+\varepsilon^2} \ ,
\end{equation}
where the parameter $\varepsilon$ satisfies
$\Omega_0\gg\varepsilon\gg\delta\omeg$. Using then the dynamical
equation (\ref{dynamics}) one obtains
\begin{eqnarray}
\tilde{S}^{(\varepsilon)}(\omega) = \frac{\hbar}{m\pi} \theta(\omega)\mathrm{Im}
  \left(\frac{1}{{\Omega}_0^2 - ({\omega-i\varepsilon})^2 -  
      i ({\omega-i\varepsilon}) \tilde{\gamma}_{\varepsilon}({\omega})} \right) \ ,
\end{eqnarray}
where $\tilde{\gamma}_{\varepsilon}(\omega)$ is the smoothened damping
kernel
\begin{equation}
\tilde{\gamma}_{\varepsilon}(\omega)= 
\int\limits_0^\infty \exp\left(\left(i\omega - \epsilon\right)t\right)\ \gamma(t)\ dt.
\end{equation}
In the case when the spectral density $\sigma$ obeys the Ohmic law,
$\tilde{\gamma}_{\varepsilon}(\omega)=\gamma_0$ is constant and we
find for the averaged $\tilde{S}(\omega)$ the expression
\begin{equation}
\tilde{S}^{(\varepsilon)}(\omega)\approx \frac{\hbar}{m\pi}\theta(\omega)
      \left(\frac{{\omega} \gamma_0}{({\Omega}_0^2 - {\omega}^2)^2  
       +({\omega} \gamma_0)^2} \right) \ .
\label{ohmictransitions}
\end{equation}
Here we choose the parameter $\varepsilon$ to be small compared to
$\gamma_0$.  In the case of an underdamped oscillator
$\Omega_0>\gamma_0/2$, the above expression can be
conveniently represented through the parameters of the corresponding
classical evolution of the collective coordinate described by equation
(\ref{expectationeq}). Hence we have
\begin{eqnarray}
X(t) =
\frac{P}{m\bar{\Omega}_0} \exp(-\bar{\gamma}_0 t) \sin(\bar{\Omega}_0 t), \quad 
 \bar{\Omega}_0=\sqrt{\Omega^2_0-\frac{\gamma_0^2}{4}}, \quad 
  \bar{\gamma}_0=\gamma_0/2 \ .
\end{eqnarray}   
With the parameters $\bar{\Omega}_0,\bar{\gamma}_0$ equation
(\ref{ohmictransitions}) takes the form
\begin{eqnarray}
\tilde{S}(\omega) =
 \theta(\omega)\ \frac{\hbar\bar{\gamma}_0}{ 2\pi m\bar{\Omega}_0} 
     \left(\frac{1}{(\omega - \bar{\Omega}_0)^2 + \bar{\gamma}_0^2} - 
     \frac{1}{(\omega + \bar{\Omega}_0)^2 + \bar{\gamma}_0^2} \right) \ , 
\label{ohmictransitions1}
\end{eqnarray}   
where we dropped the index $\varepsilon$.  In a strongly underdamped
regime $\Omega_0\gg \gamma_0/2$ the transition strength
distribution (\ref{ohmictransitions1}) has a maximum at the frequency
$\omega\approx \Omega_0\approx \bar{\Omega}_0$ of the collective
motion, and the width of the distribution is controlled by $\gamma_0$,
see fig.~(\ref{figure}). On the other hand, in the overdamped regime
$\Omega_0< \gamma_0/2$ the maximum is shifted away from
$\Omega_0$ and the distribution becomes very broad i.e., there are no
pronounced collective excitations.

\subsection{Transition strengths for general couplings}

\begin{figure}[htb]
\begin{center}{ 
\includegraphics[scale=1.0]{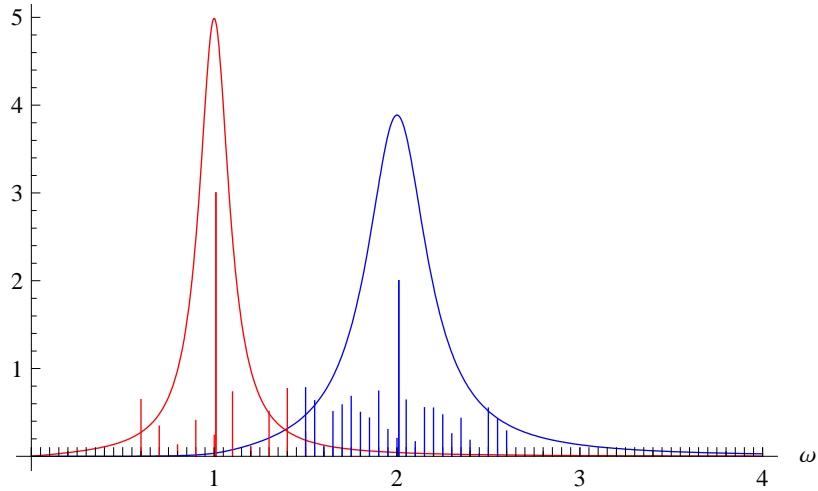}
}\end{center} 
\caption{The dimensionless functions $\left( {\pi
m\bar{\Omega}^2_0}/{\hbar}\right) \tilde{S}(\omega)$,  $\Bigl({\pi m
\bar{\Omega}^{3/2}_0}/{\hbar \sqrt{2}}\Bigl)^2  \tilde{S}^{(2)}(\omega)$
 are  plotted  on the left-hand side (red) and the right-hand side (blue)
for  the parameters $\bar{\Omega}_0=1$, $\bar{\gamma}_0=0.1$. The spikes
at the bottom of the figure schematically depict the states which are
coupled to the ground state through the operator $\hat{X}$ and
$\hat{X}^2$, respectively.} 
\label{figure}
\end{figure}

We notice that the function $\tilde{S}(\omega)$, derived in the
previous section, has only one maximum at a frequency near $\Omega_0$.
Translating this into the energy domain one concludes that the
collective excitations show up only for the first energy level
$E_1=E_0+\Omega_0\hbar$ of the damped harmonic oscillator, rather than
for all energies $E_n=E_0+n\Omega_0\hbar$.  This is directly connected
with the choice of the coupling $A(X)$ and the linear nature of our
model, since in a harmonic oscillator the transitions induced by
$\hat{X}$ only happen between neighboring states. Let us show that for
a more general choice of the coupling $A(X)$ other collective
excitations show up at energies $E_n$, $n>1$ of the collective
oscillator mode. For the sake of simplicity of exposition we will
first consider the case $A(X)= X^2$ and then comment on the general
case.  We thus consider the time correlator
\begin{eqnarray}
S^{(2)}(t):=\vacl \hat{X}^2(t) \hat{X}^2(0)\vacr - \vacl\hat{X}^2(0)\vacr^2,
\end{eqnarray}
whose Fourier transform keeps information about the transition
strengths induced by the operator $\hat{X}^2$,
\begin{eqnarray}
\tilde{S}^{(2)}(\omega)&:=& \frac{1}{2\pi} \int\limits_{-\infty}^{\infty} dt\ 
         e^{i\omega t} S^{(2)}(t)\nonumber\\
&\ = &
\sum_{m\neq 0}^{2N} |\vacl \hat{X}^2| m \rangle|^2 \delta\left(\omega - 
          \frac{E_m-E_0}{\hbar}\right) \ .
\end{eqnarray}
It is easy to show that this quantity can be expressed in terms of
$\tilde{S}(\omega)$. Indeed, separating the collective mode into
annihilation and creation parts,
\begin{eqnarray}
\hat{X}(t) = \hat{X}^+(t) + \hat{X}^-(t), \qquad  \hat{X}^+(t)\vacr=0 \ , \quad 
\vacl \hat{X}^-(t) = 0 \ ,
\end{eqnarray}
and using their commutation relation leads to
\begin{eqnarray}
S^{(2)}(t)=\vacl \hat{X}^2(t) \hat{X}^2(0)\vacr -\vacl\hat{X}^2(0)\vacr^2 = 2 S^2(t) \ .
\end{eqnarray}
This immediately implies
\begin{eqnarray}
\tilde{S}^{(2)}(\omega) =
\frac{1}{ \pi} \int\limits_{-\infty}^{\infty} dt\,\exp(i\omega t) S^2(t)
=2\int\limits_{-\infty}^{\infty} \tilde{S}(\omega') \tilde{S}(\omega - \omega') d\omega'.
\end{eqnarray}
Using then equation~(\ref{s1def}), we obtain 
\begin{eqnarray}
\tilde{S}^{(2)}(\omega)
=-8\int\limits_{0}^{\omega} \tilde{S}_1(\omega') \tilde{S}_1(\omega - \omega') d\omega'.
\end{eqnarray}
If $\sigma$ obeys an Ohmic law and if we are in the underdamped
regime, the last expression takes the form
\begin{eqnarray}
\tilde{S}^{(2)}(\omega) &=&
2\left(\frac{\hbar\bar{\gamma}}{2\pi m\bar{\Omega}_0}\right)^2 
\int\limits_0^\omega    \left(\frac{1}{(\omega' + \bar{\Omega}_0)^2 + \bar{\gamma}_0^2} - 
  \frac{1}{(\omega' - \bar{\Omega}_0)^2 + \bar{\gamma}_0^2} \right)\nonumber \\
&& \qquad \left(\frac{1}{(\omega-\omega' + \bar{\Omega}_0)^2 + \bar{\gamma}_0^2} 
 - \frac{1}{(\omega-\omega' - \bar{\Omega}_0)^2 + \bar{\gamma}_0^2} \right) d\omega'.
\end{eqnarray}  
The function $\tilde{S}^{(2)}(\omega)$ is depicted in figure (\ref{figure}).   
For $\Omega_0\gg 2\gamma_0$ (i.e., strongly underdamped regime) one
can clearly see a spike in the vicinity of the oscillator frequency $2
\Omega_0$ with the width of the spike being twice the width of
$\tilde{S}(\omega)$ for the same parameters $\gamma_0$, $\Omega_0$.

It is straightforward to generalize the above discussion to generic
observables of the form $A(\hat{X})$ using the Taylor expansion
\begin{equation}
 A(\hat{X})=\sum_{n=0}^\infty\alpha_n \hat{X}^n.
\end{equation}
After substituting this into the definition of the time correlator,
and applying Wick's thorem to the products of $X(t)$ we obtain
\begin{equation}
 S_A(t)=\vacl A(\hat{X}(t))A(\hat{X})\vacr= \sum_{n=0}^\infty\beta_n S^n(t),
\end{equation}
where $ \beta_n$ are some coeficients having dimension of inverse
length in power $2n$. Taking now the Fourier transform from both sides
of this expression we obtain for the spectral function
\begin{equation}
\tilde{S}_A(\omega)= \frac{1}{2\pi}\int\limits_{-\infty}^{\infty} dt\,e^{it\omega} S_A(t)
               = \sum_{n=0}^\infty\beta_n 
            \underbrace{\tilde{S}(\omega)*\tilde{S}(\omega)*\dots *\tilde{S}(\omega)}_{n \ {\rm times}} \ ,
\label{specfgeneral}
\end{equation}
where the symbol $*$ stands for the convolution. It is quickly seen
that, in the underdamped regime, the $n$-th term of the sum
(\ref{specfgeneral}) has a maximum at the vicinity of $n\Omega_0$ with
a width given by $n\gamma$.

\section{Conclusions}

We studied collective behavior in an integrable model consisting of
two coupled chains of harmonic oscillators.  We chose the rescaled
difference of the center of mass modes of the chains as a collective
coordinate $X$, and mapped our system onto a model of Caldeira-Leggett
type.  The resemblance with the well-known Caldeira-Leggett model
provides an intuitive physical picture of the energy exchange between
the collective coordinate and the remaining degrees of freedom playing
the role of the internal bath. As a result, the dynamics of the
collective mode is described by the damped harmonic oscillator
equation. We then relate this dynamical equation to the problem of the
existence of collective quantum excitations in the spectrum of the
corresponding quantum Hamiltonian.  These collective excitations are
probed through the transition strengths induced by observables
$A(\hat{X})$, depending on the collective coordinate. As we show, for
the dynamically underdamped regime the spikes in the distribution of
the transition strengths appear precisely at the energies $E_n = E_0
+n\hbar\Omega_0$ ($E_0 = \mathrm{ground\ state\ energy}$) of the
quantized collective harmonic oscillator, while the width of the
spikes is controled by the damping coeficient $\gamma_0$ of the
corresponding dynamical problem.  It is worth mentioning that based on
fluctuation-dissipation type of arguments we can extend the present
approach to any Hamiltonian system with quadratic interactions.

One of the important features of our model is the freedom of choice
for the collective coordinate. Note that our definition of $X$ in a
technical sense was somewhat arbitrary. In principle, we could take
any linear combination $Y=\sum_{i=1}^N \left(C_i x_i + \bar{C}_i \bar{x}_i\right)$
as a collective coordinate, and implement the same type of mapping
procedure (as in the case of $X$) onto the model of Caldeira-Legget
type.  We would get then precisely the same equation of motion for
$Y(t)$, but with a different collective frequency $\Omega_0$ and
damping kernel $\gamma(t)$. Not every choice for $Y$ would be, of
course, appropriate in order to regard it as a collective coordinate.
If, for instance, the resulting dynamics becomes overdamped, no clear
spikes will be visible at the corresponding spectral function.  On the
other hand, it seems that there exists no ``unique" choice for the
collective coordinate. This means the parameters $\Omega_0$,
$\gamma_0$ are not intrinsic properties of the considered integrable
model but are rather affected by the definition of the collective
coordinate.
It would be of a great interest to see whether and in what form the
above ``semiclassical" connection between the classical dynamics of a
collective mode and collective excitations of the corresponding
quantum problem can be extended to a more general class of
non-integrable systems.  It is clear that some substantial differences
with an integrable case must arrise when the dynamics of the system
becomes chaotic.

\section*{Acknowledgement}

We thank Heiner Kohler for fruitful discussions.  We acknowledge
support from Deutsche Forschungsgemeinschaft within
Sonderforschungsbereich Transregio 12 ``Symmetries and Universality
in Mesoscopic Systems''.


\end{document}